\documentclass[twocolumn]{jpsj3} 
\usepackage{color}
\usepackage{ulem}

\title{Superconductivity in Noncentrosymmetric Iridium Silicide Li$_2$IrSi$_3$}

\author{
Sunseng Pyon$^{1,2}$\thanks{Present address: Department of Applied Physics, The University of Tokyo, Tokyo 113-8656, Japan}, 
Kazutaka Kudo$^{1,2}$, 
Jun-ichi Matsumura$^{1,2}$, 
Hiroyuki Ishii$^{1,2}$, \\
Genta Matsuo$^{1,2}$, 
Minoru Nohara$^{1,2}$\thanks{nohara@science.okayama-u.ac.jp}, 
Hajime Hojo$^3$,
Kengo Oka$^3$,
Masaki Azuma$^3$,\\
V. Ovidiu Garlea$^4$,
Katsuaki Kodama$^5$, and
Shin-ichi Shamoto$^5$
}

\inst{
$^{1}$Department of Physics, Okayama University, Okayama 700-8530, Japan\\
$^{2}$Research Center of New Functional Materials for Energy Production, Storage and Transport, Okayama University, Okayama 700-8530, Japan\\
$^{3}$Materials and Structures Laboratory, Tokyo Institute of Technology, 4259 Nagatsuta, Midori-ku, Yokohama 226-8503, Japan\\
$^{4}$Quantum Condensed Matter Division, Oak Ridge National Laboratory, Oak Ridge, Tennessee 37831, USA\\
$^{5}$Materials Science Research Unit, Quantum Beam Science Center, Sector of Nuclear Science Research, Japan Atomic Energy Agency, Tokai, Ibaraki 319-1195, Japan
}

\abst{
The effects of lithium absorption on the crystal structure and electronic properties of IrSi$_3$, a binary silicide with a noncentrosymmetric crystal structure, were studied. X-ray and neutron diffraction experiments revealed that hexagonal IrSi$_3$ (space group $P6_{3}mc$) transforms into trigonal Li$_2$IrSi$_3$ (space group $P31c$) upon lithium absorption. The structure of Li$_2$IrSi$_3$ is found to consist of a planar kagome network of silicon atoms with Li and Ir spaced at unequal distances between the kagome layers, resulting in a polar structure along the $c$-axis.  Li$_2$IrSi$_3$ exhibited type-II superconductivity with a transition temperature $T_c$ of 3.8 K, displaying a structure type that no previous superconductors have been reported to have. 
}

\begin{document}
\maketitle

Silicides constitute a wide variety of superconducting materials. 
Depending on the various silicon networks, such materials may exhibit superconductivity at relatively high transition temperatures $T_c$. 
Prominent examples of such superconductors include 
clathrate Ba$_8$Si$_{46}$ ($T_c$ = 8 K) \cite{ref_1,ref_2}, 
honeycomb CaSi$_2$ ($T_c$ = 14 K at 12 GPa)  \cite{ref_3} and BaSi$_2$ ($T_c$ = 6.8 K) \cite{ref_4}, 
hyper-honeycomb SrSi$_2$ ($T_c$ =  3.1 K) \cite{ref_5}, 
and pyrochlore Ca(Al,Si)$_2$ ($T_c$ = 2.6 K) \cite{ref_6}. 
Transition metal silicides also exhibit superconductivity \cite{ref_7,ref_8}, as best exemplified by the superconductivity in 
V$_3$Si ($T_c$ =  17 K) \cite{ref_7}.
Most silicides, including the aforementioned ones, crystallize with centrosymmetric crystal structures and exhibit conventional $s$-wave superconductivity.
In contrast, some silicides are known to crystallize with polar noncentrosymmetric crystal structures. 
Such crystals lack spatial inversion symmetry, and because of the antisymmetric spin-orbit coupling (ASOC), the parity of the superconducting wave function can be mixed, and the pairing state cannot be classified as either singlet or triplet and may be of mixed character \cite{ref_9,ref_10,ref_11}. 
This may result in novel superconducting properties, for instance, an upper critical field $H_{c2}$ exceeding the Pauli limiting field. 
Prominent examples of such superconductors include heavy fermion superconductors with polar noncentrosymmetric structures such as 
CePt$_3$Si ($T_c$ = 0.75 K and $H_{c2}$ = 5 T) \cite{ref_12} and 
CeRhSi$_3$ ($T_c$ = 1.1 K and $H_{c2}$ = 30 T) \cite{ref_13}. 
In contrast, superconductors of transition metal silicides with similar noncentrosymmetric crystal structures but without cerium exhibit rather conventional behaviors, as reported for 
CaIrSi$_3$ ($T_c$ = 3.6 K and $H_{c2}$ = 0.27 T) \cite{ref_14}, 
CaPtSi$_3$ ($T_c$ = 2.3 K and $H_{c2}$ = 0.15 T)  \cite{ref_14}, BaPtSi$_3$ ($T_c$ = 2.25 K) \cite{ref_15}, and
SrAuSi$_3$ ($T_c$ = 1.54 K and $H_{c2}$ = 0.22 T) \cite{ref_16}, 
although these compounds consist of heavy transition metal elements such as Ir, Pt, and Au that are subjected to strong ASOC. 
The observation of unconventional superconductivity in transition metal compounds with noncentrosymmetric crystal structures has been limited to platinum boride Li$_2$Pt$_3$B, in which spin triplet superconductivity has been confirmed by nuclear magnetic resonance (NMR) measurements \cite{ref_17}. 
Thus, further exploration of noncentrosymmetric compounds with a different class of silicon networks should help to identify unconventional superconductivity in heavy transition metal silicides.

In this paper, we report that superconductivity emerges at $T_{c}$ = 3.8 K upon the lithium intercalation of noncentrosymmetric IrSi$_3$. 
IrSi$_3$ crystallizes in a hexagonal structure with the space group $P6_{3}mc$ ($\sharp$ 186) \cite{ref_18}. 
This structure consists of infinite, planar, four-connected layers of silicon atoms, i.e., a distorted kagome network, perpendicular to the $c$-axis, with each iridium atom spaced at unequal distances between the adjacent silicon layers \cite{ref_18}; therefore, the crystal structure is strongly polar noncentrosymmetric along the $c$-axis, as shown in Fig.~1(a). 
We demonstrate that hexagonal IrSi$_3$ absorbs lithium to form Li$_2$IrSi$_3$ with a noncentrosymmetric trigonal structure with the space group $P31c$ ($\sharp$ 159). Electrical and magnetic measurements reveal that Li$_2$IrSi$_3$ exhibits superconductivity at 3.8 K, and specific heat and $H_{c2}$ measurements indicate conventional superconductivity. The absence of unconventional superconductivity is likely due to the weakened inversion symmetry breaking by lithium absorption.

Polycrystalline samples of Li$_x$IrSi$_3$ were prepared by the arc-melting method.  
First, we synthesized IrSi$_3$, the precursor, by arc-melting Ir and Si powders under Ar atmosphere. 
Then, Li and IrSi$_3$ with a ratio of $x$:1 (0 $\leq$ $x$ $\leq$ 3) were arc-melted.
An almost single-phase Li$_x$IrSi$_3$ was obtained for $x$ $\simeq$ 2.
The structure of the samples was examined by electron diffraction, powder X-ray diffraction, and powder neutron diffraction measurements. 
The magnetization $M$ was measured from 1.8 to 4.5 K under a magnetic field of 10 Oe using a SQUID magnetometer (Quantum Design MPMS).
The electrical resistivity $\rho$ and specific heat $C$ were measured using a Quantum Design PPMS.
Electron diffraction patterns were taken using a transmission electron microscope (JEOL JEM-2100F).  Synchrotron X-ray powder diffraction (SXRD) data were collected with a large Debye-Scherrer camera installed at the BL02B2 beamline of SPring-8 \cite{ref_19}, 
and analyzed using the RIETAN-FP program \cite{ref_20}.
We performed neutron powder diffraction measurements of Li$_2$IrSi$_3$ (weight: 1.047 g) using the HB-2A high-resolution powder diffractometer (at 295 K with a wavelength of 1.5395 {\AA} and collimations of 12'-open-6') of the High Flux Isotope Reactor (HFIR) at the Oak Ridge National Laboratory (ORNL). 
The structural parameters were refined by the Rietveld method using the GSAS program \cite{ref_21,ref_22}.

Figures 2(a) and 2(b) show the electron diffraction patterns of the Li$_x$IrSi$_3$ ($x \simeq 2$) phase taken along the [001] and [1$\bar{1}$0] zone axes, respectively. These patterns show the Li$_x$IrSi$_3$ phase to have a trigonal or hexagonal unit cell with lattice parameters of $a \simeq$ 5.0 {\AA} and $c \simeq$ 7.8 {\AA}, and the reflections are indexed accordingly. 
There are systematic absences in the reflections: the reflection conditions are 00$l$ with $l = 2n$ and $hhl$ with $l = 2n$, 
confirming that the space group is either $P31c$ ($\sharp$ 159), $P\bar{3}1c$ ($\sharp$ 163), $P6_3mc$ ($\sharp$ 186), $P\bar{6}2c$ ($\sharp$ 190), or $P6_3/mmc$ ($\sharp$ 194). 
Because of the similarities in composition and lattice parameters, the PuRu$_3$B$_2$ structure with a Ru kagome network was chosen as the initial structural model. 
Rietveld analysis of SXRD data gave satisfactory low $R$-factors only for $P31c$ ($\sharp$  159) with a distorted Si kagome network, as shown in Fig.~2(c). 
Finally, the Rietveld refinements of neutron diffraction data were performed on this model, as shown in Fig.~2(d). 
The refined lattice parameters $a$ and $c$, atomic positions, and occupancy of Li are summarized in Table I. 
A lithium vacancy exists, and therefore $x$ = 1.8 for Li$_x$IrSi$_3$. 
Here, we use the nominal chemical formula Li$_2$IrSi$_3$ for Li$_x$IrSi$_3$ with $x$ = 1.8.

Figure~1(b) shows the crystal structure of Li$_2$IrSi$_3$.
The structure consists of an infinite-planar kagome network of silicon atoms perpendicular to the $c$-axis; therefore, the kagome network of IrSi$_3$ was preserved in Li$_2$IrSi$_3$. 
Each iridium atom was spaced at unequal distances between the two silicon layers, making the crystal structure polar noncentrosymmetric along the $c$-axis. 
The displacement from the symmetrical equidistant planes is $\Delta z/c \simeq 0.007$ along the $c$-axis, which is one order of magnitude smaller than that of $\Delta z/c \simeq 0.083$ for IrSi$_3$, suggesting that the polar symmetry breaking is considerably weakened in Li$_2$IrSi$_3$. 
Lithium atoms are located out of the iridium planes, as shown in Fig.~1(d). 
The distortion of the silicon kagome network is also weakened in Li$_{2}$IrSi$_3$; 
bond alternation in the silicon kagome network of 2.107 and 2.536 {\AA} for IrSi$_3$ is reduced to 2.402 and 2.616 {\AA} for Li$_{2}$IrSi$_3$, respectively. 
Another feature of interest is that the corner-sharing Si triangles in the kagome network rotate alternately by $\simeq$ 2.4$^{\circ}$ for Li$_2$IrSi$_3$, as shown in Fig. 1(c), while there exists no rotation for IrSi$_3$.  The alternate rotation is an additional source of broken spatial inversion symmetry that is characteristic of Li$_2$IrSi$_3$ with the symmetry of the space group $P31c$.

\begin{figure}[t]
\begin{center}
\includegraphics[width=8.5cm]{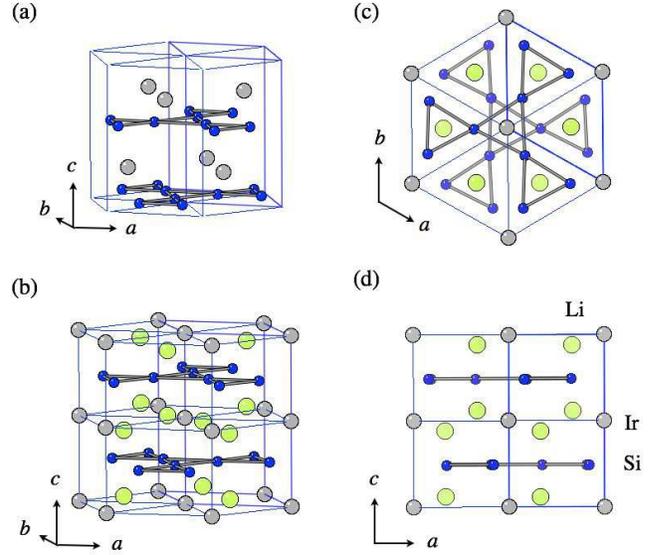}
\caption{
(Color online) 
Crystal structures of (a) IrSi$_3$ (hexagonal, space group $P6_{3}mc$) and (b) Li$_2$IrSi$_3$ (trigonal, space group $P31c$). Blue circles, yellow circles, and gray circles denote Si, Li, and Ir, respectively. (c) and (d) show the top and side views of the crystal structure of Li$_2$IrSi$_3$. }
\end{center}
\end{figure}

\begin{figure}
\begin{center}
\includegraphics[width=8cm]{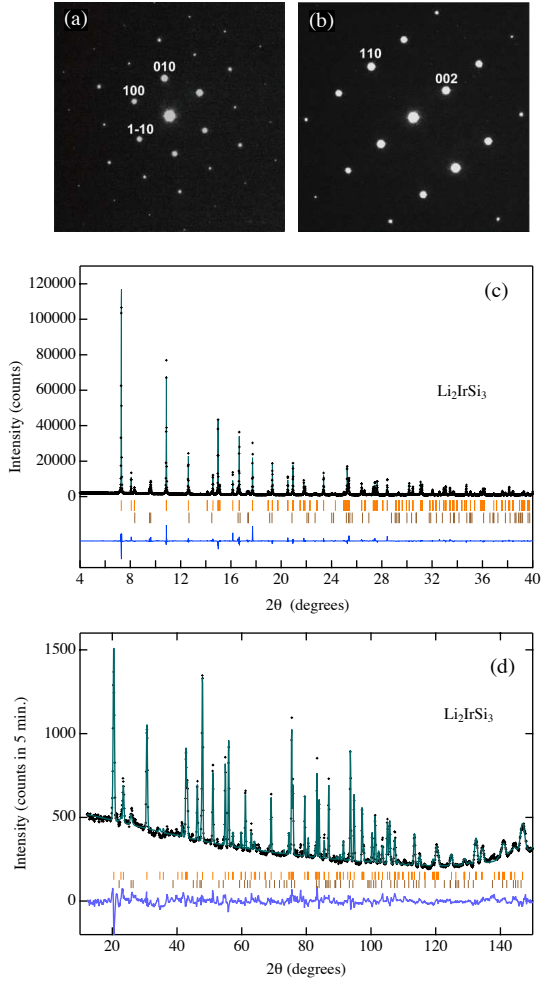}
\caption{
(Color online)
(a), (b) Electron diffraction patterns of Li$_2$IrSi$_3$ taken with the incidence of a parallel beam along the [001] and [1$\bar{1}$0] zone axes.
(c) Synchrotron powder X-ray diffraction pattern measured at room temperature and its Rietveld refinement for Li$_2$IrSi$_3$. 
The X-ray wavelength is 0.54910 {\AA}.
(d) Neutron powder diffraction pattern collected at HB-2A of HFIR at 295 K and its Rietveld refinement for Li$_2$IrSi$_3$. 
The neutron wavelength is 1.5395 {\AA}.
The vertical ticks in (c) and (d) indicate the Bragg reflection positions calculated for Li$_2$IrSi$_3$ (orange) and IrSi$_3$ (brown).
}
\end{center}
\end{figure}

\begin{table}
\caption{
Crystallographic parameters of Li$_2$IrSi$_3$ as determined by powder neutron diffraction at 300 K. 
$R_{wp}$ $=$ 4.79\% and $R_{p}$ $=$ 3.78\%. 
}
\begin{tabular}{cccccc}
\hline
\hline
\multicolumn{5}{c}{Li$_2$IrSi$_3$}\\
\hline
\multicolumn{5}{l}{Trigonal with a space group $P31c$ ($\sharp$ 159).} \\
\multicolumn{5}{l}{$a$ $=$ 5.0139(2) \AA, $c$ = 7.8397(3) \AA.} \\
\hline
\multicolumn{5}{c}{atomic positions}\\
site & occupancy & $x/a$ & $y/b$ & $z/c$ & $B$ (\AA$^2$)\\
2a & Ir 1.0   &   0.00  &  0.00 &  0.00 & 0.70(4)\\
2b &  Li 0.89(4)   &   1/3 & 2/3  & 0.0589(10)  & 0.35(24)\\
2b &  Li 0.89   &   2/3 & 1/3  & $-$0.0589 & 0.35\\
6c &  Si 1.0   &  0.81424 & 0.65238  & 0.24317  & 0.40(5)\\
\hline
\hline
\end{tabular}
\end{table}

Figure 3 shows the temperature dependences of the electrical resistivity $\rho$ for polycrystalline IrSi$_3$ and Li$_2$IrSi$_3$. 
Both compounds exhibited metallic behavior. 
It was found that the Li absorption of IrSi$_3$ reduces the residual resistivity $\rho$(0) and thus increases the residual resistivity ratio (RRR); $\rho$(0) = 9.8 $\mu\Omega$cm and RRR $\sim$ 11 in Li$_2$IrSi$_3$, whereas $\rho$(0) = 234 $\mu\Omega$cm and RRR $\sim$ 1.3 in IrSi$_3$. 
The marked reduction in $\rho$(0) suggests that the absorbed lithium atoms donate a substantial number of charge carriers to Li$_2$IrSi$_3$.  
The decrease in resistivity can be seen at low temperatures in Fig.~3, suggesting the occurrence of superconductivity in Li$_2$IrSi$_3$. 
No superconductivity was observed down to 2 K in IrSi$_3$, as shown in Fig.~3.

\begin{figure}
\begin{center}
\includegraphics[width=5.5cm]{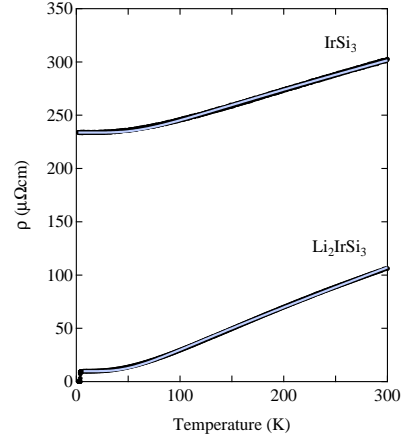}
\caption{
(Color online) 
Temperature dependences of electrical resistivity $\rho$ for polycrystalline IrSi$_3$ and Li$_2$IrSi$_3$ samples. The solid curves represent the fitting results for the Bloch-Gr\"{u}neisen model. 
}
\end{center}
\end{figure}
\begin{figure}
\begin{center}
\includegraphics[width=6cm]{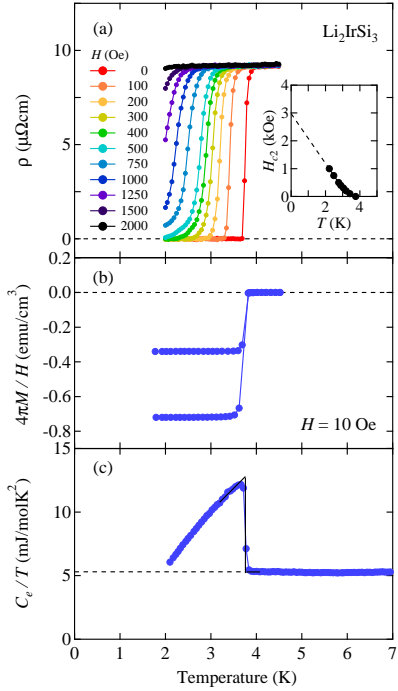}
\caption{
(Color online) 
(a) Temperature dependence of electric resistivity $\rho$ for Li$_{2}$IrSi$_{3}$ at magnetic fields $H$ up to 2000 Oe. The inset shows the temperature dependence of the upper critical field $H_{c2}$ deduced from $\rho$ measurements. The broken line represents the linear extrapolation of $H_{c2}(T)$. 
(b) Temperature dependence of dc magnetization $M$ measured at $H$ = 10 Oe for Li$_{2}$IrSi$_{3}$ under conditions of zero-field cooling and field cooling. 
(c) Temperature dependence of the electronic specific heat divided by temperature, $C_e/T$, for Li$_{2}$IrSi$_{3}$. $C_e$ was determined by subtracting the phonon contribution $\beta T^3$ from the total specific heat $C$, as shown in Fig. 5. The broken line corresponds to the electronic specific heat coefficient $\gamma$. The solid line represents an ideal jump at $T_c$, assuming entropy conservation at the transition. 
}
\end{center}
\end{figure}

Figure 4(a) shows the low-temperature data for the electrical resistivity $\rho$ of Li$_2$IrSi$_3$. 
The resistivity decreases sharply below 3.9 K and becomes zero at 3.7 K. 
The 10--90{\%} transition width was 0.1 K. 
The bulk superconductivity of Li$_2$IrSi$_3$ was evidenced by the temperature dependences of the magnetization $M$ and the specific heat $C$, as shown in Figs.~4(b) and 4(c), respectively. 
$M$ exhibited a diamagnetic behavior below 3.8 K. 
The shielding and flux exclusion signals correspond to 72 and 34{\%} of perfect diamagnetism, respectively. 
The electronic specific heat $C_e$ exhibited a clear jump at the superconducting transition. 
We estimated the superconducting transition temperature $T_c$ to be 3.75 K  and the specific heat hump $\Delta C/T_c$ to be 7.4 mJ/(mol K$^2$) by assuming the standard entropy conservation at $T_c$.

As shown in Fig.~4(a), $T_{c}$ decreases with increasing magnetic field and decreases below 2 K at 0.2 T. 
The magnetic field dependence demonstrates that Li$_2$IrSi$_3$ is a type-II superconductor. 
The inset of Fig.~4(a) shows the temperature dependence of the upper critical field $H_{c2}$ as determined from the midpoint of the resistive transition. 
$H_{c2}(T)$ exhibited an upward curvature with a visible change in the slope $dH_{c2}/dT$ near $T_{\rm c}$ = 3.8 K in the low-magnetic-field region. 
Thus, we tentatively estimated $H_{c2}(0)$ to be 0.3 T from a linear extrapolation of the data at low temperatures between 2.2 and 2.8 K.
The Ginzburg--Landau coherence length $\xi_0$ was estimated to be  330 {\AA} from $\xi_0$ = $(\Phi_0/[2\pi H_{c2}(0)])^{1/2}$, where $\Phi_0$ is the magnetic flux quantum. 
The small $H_{c2}$, which is comparable to those reported for the typical noncentrosymmetric silicide superconductors 
CaIrSi$_3$ \cite{ref_14}, CaPtSi$_3$ \cite{ref_14}, and SrAuSi$_3$ \cite{ref_16}, 
suggests a conventional superconducting state of Li$_2$IrSi$_3$. 
No spin triplet mixing can be anticipated because $H_{c2}(0)$ is considerably smaller than that expected from the Pauli limiting field, 
$H_{c2}^{\rm P}(0) = 1.84 T_{c}$ = 6.9 T, for the BCS weak coupling limit. 
Despite the low $H_{c2}(0)$, the upward curvature of $H_{c2}(T)$ may be interesting because it suggests a multiband/multigap superconductivity in Li$_2$IrSi$_3$. 
Similar behavior has been observed in MgB$_2$ \cite{ref_23,ref_24}, 
YNi$_2$B$_2$C \cite{ref_25,ref_26}, LaFeAsO$_{0.89}$F$_{0.11}$\cite{ref_27}, and SrPtAs \cite{ref_28}.

The absence and emergence of superconductivity in IrSi$_3$ and Li$_2$IrSi$_3$, respectively, can partly be explained by the difference in the electronic density of states (DOS) at the Fermi level between the two compounds; 
the specific heat data exhibit a considerable increase in the DOS at the Fermi level with the lithium absorption of Li$_2$IrSi$_3$. 
The standard analysis of the normal-state specific heat, shown in Fig.~5, yielded the electronic specific heat coefficient $\gamma$ = 0.73 mJ/(mol K$^2$) and Debye temperature $\Theta_{\rm D}$ = 516 K for IrSi$_3$, and $\gamma$ = 5.3 mJ/(mol K$^2$) and $\Theta_{\rm D}$ = 484 K for Li$_2$IrSi$_3$. 
Electronic $\gamma$ exhibits a sevenfold increase, and phononic $\Theta_{\rm D}$ shows a small change from IrSi$_3$ to Li$_2$IrSi$_3$, 
suggesting the major role of the increased $\gamma$ and thus the electronic DOS at the Fermi level in the emergence of superconductivity. 
Using this estimated $\gamma$, we estimated $\Delta C/\gamma T_c$  to be 1.41 for Li$_2$IrSi$_3$, which is comparable to that of the BCS weak-coupling limit ($\Delta C/\gamma T_c$ = 1.43), suggesting that Li$_2$IrSi$_3$ is a conventional phonon-mediated superconductor. 

The presence of electron-phonon interactions can be anticipated from the temperature-dependent resistivity ($\propto T^5$ at low temperatures) for both IrSi$_3$ and Li$_2$IrSi$_3$. 
As shown in Fig.~3, the $\rho(T)$ data of both compounds can be well fitted using the phonon-assisted Bloch-Gr\"{u}neisen formula: 
\begin{equation}
\frac{1}{\rho(T)} = \frac{1}{\rho_{\rm i}} + \frac{1}{\rho_{\rm s}}, 
\end{equation}
\begin{equation}
\rho_{\rm i} = \rho(0) + A \left(\frac{T}{\Theta_{\rm D}^{*}}\right)^5\int_0^\frac{\Theta_{\rm D}^{*}}{T} \frac{x^5}{(e^x -1)(1-e^{-x})}dx, 
\end{equation}
where $\rho_{\rm s}$ is the saturation resistivity; $\Theta_{\rm D}^{*}$, the transport Debye temperature; and $A$, a constant depending on the material. 
The least-squares fitting yielded $\Theta_{\rm D}^{*}$ = 397 and 330 K for IrSi$_3$ and Li$_2$IrSi$_3$, respectively. 
The difference in $\Theta_{\rm D}^{*}$ is small, consistent with the thermally estimated $\Theta_{\rm D}$, indicating that the difference between the phonon spectra of IrSi$_3$ and Li$_2$IrSi$_3$ is small. This in turn supports the importance of the large DOS at the Fermi level for the occurrence of superconductivity.

\begin{figure}
\begin{center}
\includegraphics[width=5.5cm]{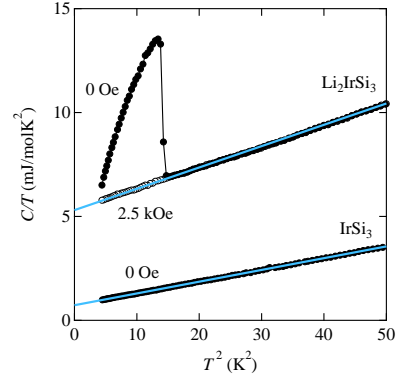}
\caption{
(Color online) 
Specific heat divided by temperature, $C/T$, as a function of $T^2$ for IrSi$_3$ at zero field, and that for Li$_2$IrSi$_2$ at zero field and a magnetic field of 2.5 kOe. The solid lines denote the fits to $C/T = \gamma + \beta T^2$, where $\gamma$ is the electronic specific heat coefficient and $\beta$ is that of the phonon contribution.
}
\end{center}
\end{figure}

Although no noticeable unique nature of superconductivity has been identified in the present thermodynamic and transport measurements of Li$_2$IrSi$_3$, 
further investigation will be worthwhile for examining a possible broken time-reversal symmetry in the superconducting state. 
A prototypical example could be SrPtAs ($T_c$ = 2.4 K and $H_{c2}$ = 0.2 T) with a locally noncentrosymmetric hexagonal structure \cite{ref_28}; $\mu$SR measurements detected the emergence of a finite spontaneous magnetic field, indicative of a broken time-reversal symmetry, in the superconducting state of SrPtAs \cite{ref_29}, for which a chiral $d$-wave superconducting state has been proposed \cite{ref_30} 
although NMR measurements suggest conventional superconductivity \cite{ref_31}. 
Another example is Mg$_{10}$Ir$_{19}$B$_{16}$ ($T_c$ $\simeq$ 5 K) with a noncentrosymmetric cubic structure \cite{ref_32}, although no sign of a spontaneous magnetic field has been detected \cite{ref_33}. 

In conclusion, we identified that Li$_2$IrSi$_3$ crystallizes to a novel trigonal structure with the space group $P31c$, which consists of an alternate stacking of planar silicon kagome planes and  Li and Ir planes. The structure is weakly noncentrosymmetric owing to the displacement of kagome planes along the $c$-axis and the staggered rotation of silicon triangles in the kagome planes. 
Superconductivity at 3.8 K was discovered in Li$_2$IrSi$_3$, although no superconductivity was detected in IrSi$_3$, which is strongly polar noncentrosymmetric. 
Weakened inversion symmetry breaking, together with the increased electronic DOS at the Fermi level, plays a key role in the superconductivity of Li$_2$IrSi$_3$. 
Silicides exhibit various silicon networks -- clathrates, honeycombs, hyper honeycombs, and pyrochlore -- and the present Li$_2$IrSi$_3$ is another example of a silicon network, namely, a kagome network. This finding will help in further understanding the relationship between noncentrosymmetric crystal structures and electronic states of silicide superconductors.

\begin{acknowledgments}
Part of this work was performed at the Advanced Science Research Center, Okayama University. 
The synchrotron radiation experiments were performed at SPring-8 with the approval of the Japan Synchrotron Radiation Research Institute (2012A1665).
Research conducted at ORNL's High Flux Isotope Reactor was sponsored by the Scientific User Facilities Division, Office of Basic Energy Sciences, US Department of Energy.
This work was partially supported by the U.S.-Japan Collaborative Program on Neutron Scattering, Grants-in-Aid for Young Scientists (B) (24740238, 26820291, and 26800180), Creative Scientific Research (26106507), Scientific Research (B) (25287094 and 26287082), and (C) (25400372) from the Japan Society for the Promotion of Science (JSPS), and the Funding Program for World-Leading Innovative R\&D on Science and Technology (FIRST Program) from JSPS.
\end{acknowledgments}

\end{document}